%% file: main.tex
\documentclass[sigconf]{acmart} 

\usepackage[acronym]{glossaries}
\usepackage{amsthm, cleveref}

\usepackage{hyperref}
\usepackage{subfig}

\usepackage{todonotes}

\theoremstyle{definition}
\newtheorem{definition}{Definition}[section]

\newacronym{CS}{CS}{Construct Space}
\newacronym{OS}{OS}{Observed Space}
\newacronym{DS}{DS}{Decision Space}
\newacronym{PS}{PS}{Potential Space}
\newacronym{WYSIWYG}{WYSIWYG}{"What You See Is What You Get"}
\newacronym{MB}{MB}{Measurement Bias}
\newacronym{LB}{LB}{Life's Bias}
\newacronym{WAE}{WAE}{"We're All Equal"}
\newacronym{WAEPS}{WAEPS}{"We're All Equal in the \gls{PS}"}
\newacronym{WAECS}{WAECS}{"We're All Equal in the \gls{CS}"}
\newacronym{STEM}{STEM}{science, technology, engineering, and mathematics}
\newacronym{WAEBI}{WAEBI}{"We're All Equal But There Is Injustice"}

\AtBeginDocument{%
  \providecommand\BibTeX{{%
    \normalfont B\kern-0.5em{\scshape i\kern-0.25em b}\kern-0.8em\TeX}}}

\copyrightyear{2021}
\acmYear{2021}
\setcopyright{acmlicensed}
\acmConference[FAccT '21]{Conference on Fairness, Accountability, and Transparency}{March 3--10, 2021}{Virtual Event, Canada}
\acmBooktitle{Conference on Fairness, Accountability, and Transparency (FAccT '21), March 3--10, 2021, Virtual Event, Canada}
\acmPrice{15.00}
\acmDOI{10.1145/3442188.3445936}
\acmISBN{978-1-4503-8309-7/21/03}

\begin{document}

\title{On the Moral Justification of Statistical Parity}

\author{Corinna Hertweck}
\email{corinna.hertweck@zhaw.ch}
\orcid{}
\affiliation{%
  \institution{Zurich University of Applied Sciences, University of Zurich}
}

\author{Christoph Heitz}
\email{christoph.heitz@zhaw.ch}
\orcid{}
\affiliation{%
  \institution{Zurich University of Applied Sciences}
}

\author{Michele Loi}
\email{michele.loi@ibme.uzh.ch}
\orcid{}
\affiliation{%
  \institution{University of Zurich}
}


\begin{abstract}
A crucial but often neglected aspect of algorithmic fairness is the question of how we justify enforcing a certain fairness metric from a moral perspective.
When fairness metrics are proposed, they are typically argued for by highlighting their mathematical properties.
Rarely are the moral assumptions beneath the metric explained.
Our aim in this paper is to consider the moral aspects associated with the statistical fairness criterion of independence (statistical parity).
To this end, we consider previous work, which discusses the two worldviews \acrfull{WYSIWYG} and \acrfull{WAE} and by doing so provides some guidance for clarifying the possible assumptions in the design of algorithms.
We present an extension of this work, which centers on morality.
The most natural moral extension is that independence needs to be fulfilled if and only if differences in predictive features (e.g. high school grades and standardized test scores are predictive of performance at university) between socio-demographic groups are caused by unjust social disparities or measurement errors.
Through two counterexamples, we demonstrate that this extension is not universally true.
This means that the question of whether independence should be used or not cannot be satisfactorily answered by only considering the justness of differences in the predictive features.
\end{abstract}




\maketitle

\input{chapters/01_introduction}
\input{chapters/02_independence_what}
\input{chapters/03_independence_when}
\input{chapters/04_enhance_wae}

\input{chapters/05_arguments}
\input{chapters/06_conclusion}

\begin{acks}
We thank our three anonymous reviewers for their helpful feedback.
This work was supported by the National Research Programme “Digital Transformation” (NRP 77) of the Swiss National Science Foundation (SNSF), grant number 187473.
\end{acks}

\bibliographystyle{ACM-Reference-Format}
\bibliography{main}

\end{document}

%% file: chapters/01_introduction.tex
\section{Introduction}

A look at current practices suggests that in order to evaluate the fairness of a given machine learning model, so-called \textit{fairness metrics} have to be computed.
However, this disregards the crucial steps which should precede the calculation of fairness metrics: discussing the moral reasons underlying the decision to select one or more fairness metrics to be enforced.
This is easily forgotten as the procedure is not as straightforward as computing statistical metrics.
It might require discussions with the stakeholders of the application and finding a compromise -- after all, there are very few cases where everyone can agree on the correct choice of fairness metrics, in particular because it has been shown that some of them are conflicting \cite{kleinberg2016inherent, fairmlbook, chouldechova2017fair}.
It is thus important to not only provide mathematical definitions of fairness metrics, but to provide some guidance on how to reason about them from a moral perspective.

One popular metric is called \textit{independence}, often referred to as \textit{statistical parity}.
While existing literature at first sight often seems to reason about independence in moral terms, a lot of the arguments are either not backed up by moral philosophy or turn out to be purely mathematical.
We note that the need for enforcing independence is rarely justified from a philosophical perspective, and that the two spaces (philosophy and mathematics) are often conflated, in part also due to terminology.
In this paper, we want to make a contribution towards resolving this ambiguity, and to highlight the relation between mathematical justifications for choosing independence and the corresponding moral significance.

We begin by defining independence mathematically in \Cref{sec:independence} and then reconstruct arguments on when independence is considered the correct fairness metric in \Cref{sec:rules-literature}.
We will show that these arguments, while at first sight appearing to hold moral value, are actually purely mathematical if taken literally.
However, since they suggest that there are moral reasons for choosing independence, we will provide a natural extension for the arguments found in the literature (\Cref{sec:moral-rules}).
We will then argue that this natural extension is not always in line with our moral intuitions about fairness in specific cases (Section \ref{sec:counterexamples}).
We conclude in Section \ref{sec:conclusion} that the question whether independence should be chosen or not is not sufficiently answered by considering the social injustices occurring from the birth of an individual to the point where their abilities are measured.

%% file: chapters/02_independence_what.tex
\section{What is independence?}\label{sec:independence}

The arguably most commonly used category of fairness metrics is referred to as \textit{group fairness} and focuses on the question whether socio-demographic groups are treated similarly or receive similar outcomes \cite{dwork2012fairness, mehrabi2019survey}.
Group fairness is tested with respect to specific socio-demographic groups, differentiated through a \textit{sensitive attribute}, which we will denote as $A$.
One of the more prominent fairness metrics falling into the category of group fairness is \textit{statistical parity} \cite{fairmlbook}.

The concept is easiest to explain for binary classification ($\hat{Y} \in {0,1}$) and two groups $A=a$ and $A \neq a$: In such a case, statistical parity requires that the probability of the predicted outcome being positive, i.e., $\hat{Y}=1$, is equal for $A=a$ and $A \neq a$.
In other words, the selection rate $P(\hat{Y}=1)$ has to be independent of the value of the sensitive attribute.
This can be expressed as $P(\hat{Y}=1|A=a)=P(\hat{Y}=1|A \neq a)$ \cite{verma2018fairness}.
This formula can be generalized for the case of not only a binary predictor, but any predictor $R$ and possibly more values for the sensitive attribute $A$: $R \perp A$ \cite{fairmlbook, mitchell2021algorithmic}.
This general proposition is referred to as the fairness criterion \textit{independence} \cite{fairmlbook}.
Independence can be found implemented in practice, in particular in the HR domain%
\footnote{
The popularity of independence in HR is attributable to the notion of \textit{disparate impact} found in the Uniform Guidelines on Employment Selection Procedures introduced by the Equal Employment Opportunity Commission (EEOC) in 1978 \cite{eeoc}.
Disparate impact means that one group is disproportionally affected by e.g. a hiring policy.
When bringing forth a disparate impact claim, the "4/5ths" rule works as a rule of thumb: It compares the share of hired people from each group.
If one group's hiring rate is less than 4/5ths of the hiring rate of the other group, then HR might be liable for disparate impact discrimination \cite{eeoc}.
This 4/5ths rule is essentially an expression of statistical parity that allows for some leeway:
Instead of enforcing the perfect equalization of hiring rates, it allows for some difference in hiring rates.
Even though this rule is just a rule of thumb, \citeauthor{raghavan2020mitigating} \cite{raghavan2020mitigating} recently showed that vendors of algorithmic hiring tools treat it as a hard constraint.
The web demo of AIF360 \cite{bellamy2018ai}, a tool built for the bias auditing of predictive models, refers to statistical parity as being fulfilled if the 4/5ths rule is fulfilled.
We thus see that through the 4/5ths rule, the notion of independence is highly relevant in practice.
}%
, and has been particularly influential in the early stages of the algorithmic fairness literature (see, e.g., \cite{calders2009building, pedreshi2008discrimination}).
Together with independence, \cite{fairmlbook} lists \textit{separation} and \textit{sufficiency} as the three fairness criteria that most fairness metrics that have been proposed in the literature are closely related to.
Separation and sufficiency gained more attention in recent years through the debate sparked by \citeauthor{machine-bias}'s investigative article \textit{Machine Bias} \cite{machine-bias}.
Subsequent publications have pointed out what is now known as the \textit{impossibility theorem}: Except for in highly constrained cases, we can only satisfy one of the three fairness criteria \cite{kleinberg2016inherent, fairmlbook, chouldechova2017fair}.
This impossibility theorem forces us to pick a specific fairness criterion to enforce or to find a trade-off between them, which raises the question of when one criteria should be chosen over the other two.
In this paper, we will focus on the moral reasons for enforcing independence.

Unsurprisingly, machine learning models trained to optimize, for example, accuracy, rarely coincidentally fulfill independence.
However, we can enforce independence through various strategies (see, e.g., \cite{calders2009building, kamishima2012fairness}).
This idea is what we will refer to with the phrase "independence should be used".
We can either enforce achieving independence fully or partially, e.g., because we want to trade off fairness with another goal such as business interests, or utilitarian moral goals such as maximizing the number of lives saved.

%% file: chapters/03_independence_when.tex
\section{When should independence be used?}\label{sec:rules-literature}

The question of when independence should be chosen over separation and sufficiency cannot be answered from a purely mathematical or technical perspective.
\citeauthor{friedler2016possibility} \cite{friedler2016possibility} propose a framework which enables its users to clarify the \textit{worldview} assumed in the context of their application.
This section will discuss \citeauthor{friedler2016possibility}'s paper and distill the implicit rules for when to apply independence.\footnote{
We discuss this particular paper as it uncovers the hidden assumptions that seem to be held when algorithmic fairness scholars advocate for independence \cite{abbasi2019fairness}.
The paper is well-known in the field and has influenced both theoretical and practical work.
On the practical side, the two opposing worldviews have been described in AIF360's web demo to guide practitioners in their choice of fairness criterion \cite{bellamy2018ai}.
Theoretical work has built on \citeauthor{friedler2016possibility}'s framework to, e.g., provide mathematical ways to interpolate the proposed opposing worldviews \cite{hacker2017continuous, zehlike2020matching}.
\citeauthor{mitchell2021algorithmic} \cite{mitchell2021algorithmic} cite \citeauthor{friedler2016possibility}'s reasoning when explaining when enforcing independence should be considered.
}
We also argue that the framework presented by \citeauthor{friedler2016possibility} is insufficient to represent the philosophical debate surrounding independence and thus propose an extension of their framework.

\subsection{Existing Framework}\label{sec:framework}

The premise of \citeauthor{friedler2016possibility}'s framework is that when a decision has to be made by using data-driven predictions, this prediction is based not on the features that we would ideally have access to, but on proxies.
This is reflected in the main result of \cite{friedler2016possibility}, which is the distinction between the following three spaces (see \Cref{fig:spaces-friedler}):
\begin{itemize}
    \item the \gls{CS}, which consists of the features that we \textbf{want} to base the decision on,
    \item the \glsfirst{OS}, which consists of the features that we \textbf{actually} base the decision on because the \gls{CS} is not observable, so the \gls{OS} is our proxy for the \gls{CS}, and
    \item the \gls{DS}, which encompasses the predictions based on the \gls{OS}.
\end{itemize}

In order to clarify the theoretical explanations, we will work with an exemplary scenario throughout this paper, which to some extent has also been used in \cite{friedler2016possibility}: hiring.
In this case, the task is to predict employee productivity and take a hiring decision based on this prediction.
We borrow the language for this example from \cite{mulligan2017uncertainty}.
A company picks whom to hire from the pool of \textit{candidates}, i.e., the people who apply for the job.
In order to make this decision, the company tries to predict who will perform best when hired.
Each candidate brings certain \textit{qualifications}, based on which the company wishes to make its choice.
However, these qualifications are not directly observable (e.g., how good they are at selling the company's product, how well they fit into the existing team, etc.).
Instead, the company only has access to noisy representations of these qualifications, which we will refer to as \textit{proxies}.
These proxies typically include the CV, the motivation letter, the impressions from the interviews etc.
As the company only has access to the qualifications through the proxies, it has to base its hiring decisions on the proxies.
In this example, the qualifications are equivalent to the unobservable \gls{CS} while the proxies represent the observable \gls{OS}.

\citeauthor{friedler2016possibility} present two opposing worldviews and advocate for being transparent about which one the prediction model adheres to.
The two opposing worldviews are:
\begin{itemize}
    \item \gls{WYSIWYG}, which assumes that there are barely any differences between the \gls{OS} and the \gls{CS}, meaning that the \gls{OS} is a good proxy for the \gls{CS}.
    This implies that observed differences between socio-demographic groups correspond to actual differences.
    In our example, that would mean that the usage of CVs, interviews etc. as the proxy for the candidates' qualifications neither harms nor benefits one group more than another.
    \item \gls{MB}, which assumes that the mapping of individuals from the \gls{CS} to the \gls{OS} introduces disparities between the socio-demographic groups, implying that differences between groups in the \gls{OS} are bigger than in the \gls{CS}.
    For the hiring example, this means that using CVs as proxies of qualifications harms one group compared to another one.
\end{itemize}

\citeauthor{friedler2016possibility} refer to the second worldview as \textit{structural bias}.
However, we will call this worldview \gls{MB} in order to distinguish it from the informal usage of the term "structural bias."

Furthermore, \citeauthor{friedler2016possibility} propose the axiom \gls{WAE}, which oftentimes aligns with the worldview \gls{MB}.
\gls{WAE} assumes "that in the construct space all groups look essentially the same" and "that there are no innate differences between groups" \cite[p. 8]{friedler2016possibility}.
If we assume that there are no innate differences in the abilities of socio-demographic groups to perform well on a job, but measure differences once we evaluate their CVs, we will see these observed differences as a result of the \gls{MB}.
Due to this conceptual closeness of \gls{WAE} and \gls{MB} the literature oftentimes presents \acrfull{WAE} and \acrfull{WYSIWYG} as the two opposing worldviews.

\subsection{Our Extension of the Framework}\label{sec:extension}

When discussing the cause of group differences in the \gls{OS}, \citeauthor{friedler2016possibility}'s framework quickly reaches its limits.
The main issue is that two different spaces are conflated and merged in the \gls{CS}.
Although \citeauthor{friedler2016possibility} recognize this, they justify not differentiating between the two by saying that a differentiation would still lead to "the same mathematical outcome" \cite[p. 8]{friedler2016possibility}.

However, we believe that it is necessary to clearly distinguish these two spaces as it increases the understanding of the decision making process and helps navigate the moral assessment.
More specifically, it clarifies at which stage the differences that we observe between groups in the \gls{OS} are introduced.
This is needed when discussing the morality of independence as independence looks at the \gls{DS}, which of course relies entirely on the \gls{OS}.
Therefore, we need terms to discuss the causes of difference in \gls{OS} if we want to morally justify enforcing independence in the \gls{DS}.

We base our extension on the work of \citeauthor{rawls2001justice} \cite{rawls2001justice} who differentiates between \textit{realized abilities} and \textit{innate potential} (or, as \citeauthor{rawls2001justice} writes, "native endowments").
Potential is innate to an individual and determined at birth.\footnote{"At birth" here should be considered as "at conception" since there is already social influence at the fetal stage  \cite{kollar2015prenatal}.}
This could, for example, be their innate intelligence or predisposition (e.g., extroversion) to develop the traits for being a good sales person.%
\footnote{"[N]ative endowments of various kinds (say, native intelligence or natural ability) are not fixed natural assets with a constant capacity. They are merely potential and cannot come to fruition apart form social conditions [...]. Educated and trained abilities are always a selection [...] from a wide range of possibilities that might have been fulfilled. Among what affects their realisation are social attitudes of encouragement and support, and institutions [...], opportunities and social position, and the influence of good and ill fortune." \cite[pp. 56-57]{rawls2001justice}}
The realized abilities represent how good job candidates actually are at making sales at the time when the company is looking to hire.
This may be influenced by early socialization in the family, the type of school they went to, the university they attended, the opportunities they were given (internships etc.) and so on.
The realized abilities is what we will keep referring to as the \gls{CS} in our extension.
We introduce a new space which represents the potential: The \acrfull{PS}.

\begin{figure}[h]
\centering
\subfloat[\citeauthor{friedler2016possibility}] {\includegraphics[width=0.7\linewidth]{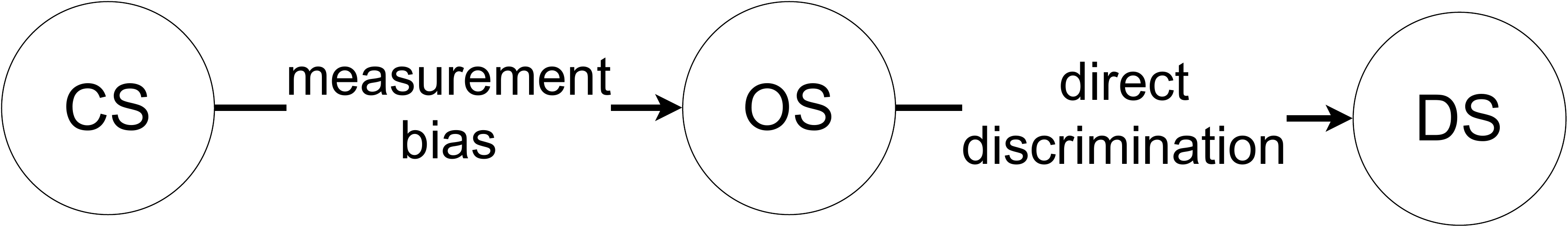}
\label{fig:spaces-friedler}}
\hfill
\subfloat[Our extension]
{\includegraphics[width=\linewidth]{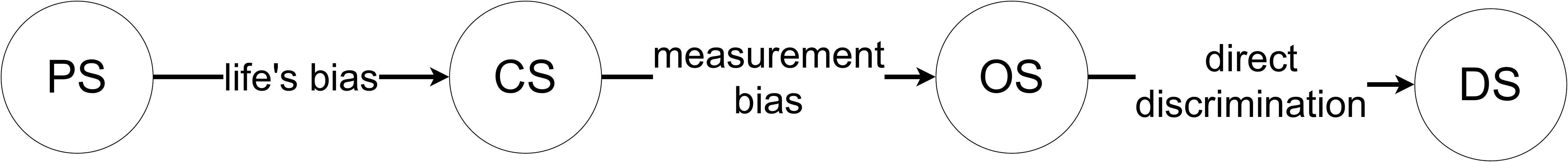}
\label{fig:spaces-extension}}\qquad
\caption{Relationship between the spaces and biases.}
\label{fig:spaces}
\end{figure}

Figure \ref{fig:spaces-extension} shows the \textit{spaces} and \textit{biases} that we differentiate in our model.
The spaces can be understood as different stages: We start with our innate potential, represented by the \gls{PS}, at birth.
Shaped by our life experiences, we realize our abilities to potentially different degrees, which is captured in the \gls{CS}.
The realized abilities are then measured in the \gls{OS}.
The \gls{OS} is used as the basis of the predictions in the \gls{DS}.%

The introduction of the \gls{PS} gives us the ability to differentiate between two types of "we're all equal", which are conflated in \citeauthor{friedler2016possibility}'s description of \gls{WAE}.
We will define them as distinct worldviews.
Note that besides assuming one of these worldviews, it is also possible to hold both views at the same time, or neither of them as they are not opposing.
\begin{itemize}
    \item \gls{WAEPS}, meaning all groups have the same innate potential.
    This means that lacking \gls{LB} (which will be defined below), all groups would have the same realized abilities.
    \item \gls{WAECS}, meaning all groups have the same realized abilities (even though it may look differently when taking measurements).
    This is the literal interpretation of "we're all equal", which implies that all groups are currently equal in their abilities.
\end{itemize}

As noted earlier, \citeauthor{friedler2016possibility} formally define \gls{WAE} as equality in the \gls{CS}, but leave the option that it may also be interpreted as equality in the \gls{PS}.
We explicitly distinguish the two worldviews as we see it as necessary for discussing independence from a philosophical perspective.

The distinction between \gls{PS} and \gls{CS} allows us to define another type of bias.
As already stated, \citeauthor{friedler2016possibility} refer to the introduction of group differences from the \gls{CS} to the \gls{OS} as "structural bias" while we refer to it as \gls{MB} since it is introduced through the act of measuring and is dependent on, e.g., availability of information or variable selection.
As seen in Figure \ref{fig:spaces-friedler}, they also term the bias introducing group differences from the \gls{OS} to the \gls{DS}: direct discrimination.
We introduce a third bias, which is the bias from the \gls{PS} to the \gls{CS}.
Inequalities, such as differences in the qualities of schools and universities, the income and connections of their parents etc., can set individuals with the same potential far apart in terms of their realized abilities.
We will refer to these inequalities as \acrfull{LB}.
We remain neutral, at this stage, on whether \gls{LB} is the same as injustice.
We notice, in passing, that if injustice exists, it may affect \gls{LB}.
For example, if people routinely act based on gender stereotypes, men and women with the same potential may end up expressing different realized abilities to a different degree.
Furthermore, if acting based on gender stereotypes is morally wrong (as it seems plausible), \gls{LB} will be unjust.
In cases like this, we shall refer to \gls{LB} as \textit{unjust} \gls{LB} for precision's sake.

\subsection{Rules Distilled From the Framework}\label{sec:rules-distilled}

We will now again consider the framework proposed by \citeauthor{friedler2016possibility} to distill rules about when and why to choose independence as the fairness measure.
The extension of \citeauthor{friedler2016possibility}'s framework presented in the previous section will be used in order to clarify the position of the paper.

In order to understand \citeauthor{friedler2016possibility}'s reasons for recommending independence, we have to introduce two key terms that appear in their proposal: Fairness and non-discrimination.
We refer to \cite{friedler2016possibility} for a mathematically precise definition, but the idea can be expressed as follows:
\begin{itemize}
    \item \textit{Fairness}: Individuals who are close in the \gls{CS} are (fairly) close in the \gls{DS}.\footnote{This corresponds to \citeauthor{dwork2012fairness}'s \textit{fairness constraint}, which requires that "similar individuals are treated similarly" \cite{dwork2012fairness}.}
    \item \textit{Non-discrimination}: The difference between groups is not (notably) increased from the \gls{CS} to the \gls{DS}.
\end{itemize}

In order to avoid confusion with the colloquial way of using these terms, we will avoid using these terms in any other way than defined by \citeauthor{friedler2016possibility}
If we do use them in the colloquial or philosophical sense, we will make this evident from here on.

We find that the paper considers the usage of independence from two perspective:
One perspective specifies the assumptions justifying the enforcement of independence (\texttt{IF [assumption], THEN independence should be used}) and the other one describes which assumptions are implied when enforcing independence is argued for (\texttt{IF independence should be used, THEN [assumption]}).
We will discuss both perspectives separately and derive a proposition summarizing them as a single rule.

\paragraph{IF [assumption], THEN independence should be used}

The first perspective asks the question what condition has to be met for suggesting the enforcement of independence.
\citeauthor{friedler2016possibility} state that "under a structural bias worldview, only group fairness mechanisms achieve non-discrimination (and individual fairness mechanisms are discriminatory)" \cite[p. 12]{friedler2016possibility}.
Note that "group fairness mechanisms" here refers to algorithms fulfilling independence (and not any group fairness metric) and that "structural bias" is what we refer to as \gls{MB}.
This statement should be interpreted to mean that if we assume \gls{MB}, only independence \textit{can} achieve non-discrimination, but it is \textit{not} a guarantee.
In fact, they present an example in which \gls{MB} is assumed, but there are differences in the \gls{CS}.
In this case, enforcing independence would still be discriminatory.
They conclude that enforcing independence only guarantees non-discrimination if \gls{WAECS} is also assumed.

This leads to the following first rule for \citeauthor{friedler2016possibility}:

\begin{proposition}\label{prop:mb-waecs-independence}
\texttt{IF there is \gls{MB}\footnote{What we mean here is "we assume that there is \gls{MB}". For brevity, however, we will simply write "there is \gls{MB}" from now on.} AND \gls{WAECS}\footnote{Again, we will write "\gls{WAECS}" to say "we assume that \gls{WAECS}".},\\
THEN independence should be used}.
\end{proposition}

\paragraph{IF independence should be used, THEN [assumption]}

\citeauthor{friedler2016possibility} claim that "under a \gls{WYSIWYG} worldview [i.e., no \gls{MB}], [...] group fairness mechanisms [i.e., independence] are unfair" \cite[p. 12]{friedler2016possibility}.
This can be translated as \texttt{IF \gls{WYSIWYG}, THEN NOT independence should be used} because we otherwise create unfairness.
Since \texttt{\gls{WYSIWYG}} is equivalent to \texttt{NOT there is \gls{MB}}, the rule can be restated as \texttt{IF NOT there is \gls{MB}, THEN NOT independence should be used}.
From that we know that independence should only be used if there is \gls{MB}.
After all, if there is no \gls{MB}, then independence should not be used.
From this we can follow the other side of the rule: \texttt{IF independence should be used, THEN there is \gls{MB}} because we otherwise create unfairness.

Further, \citeauthor{friedler2016possibility} write that when "the goal is to bring this difference [between groups in the \gls{DS}] close to zero, the assumption is that groups should, as a whole, receive similar outcomes. This reflects an underlying assumption of the we’re all equal axiom so that similar group outcomes will be non-discriminatory" \cite[p. 14]{friedler2016possibility}
Using independence thus also reflects the assumption that \gls{WAECS} holds, so \texttt{IF independence should be used, THEN \gls{WAECS}}.

We follow the second part of the rule as follows:
\begin{proposition}\label{prop:independence-mb-waecs}
\texttt{IF independence should be used, THEN there is \gls{MB} AND \gls{WAECS}}.
\end{proposition}

\paragraph{Merging the rules}

\begin{figure}[h]
    \centering
    \includegraphics[width=0.9\linewidth]{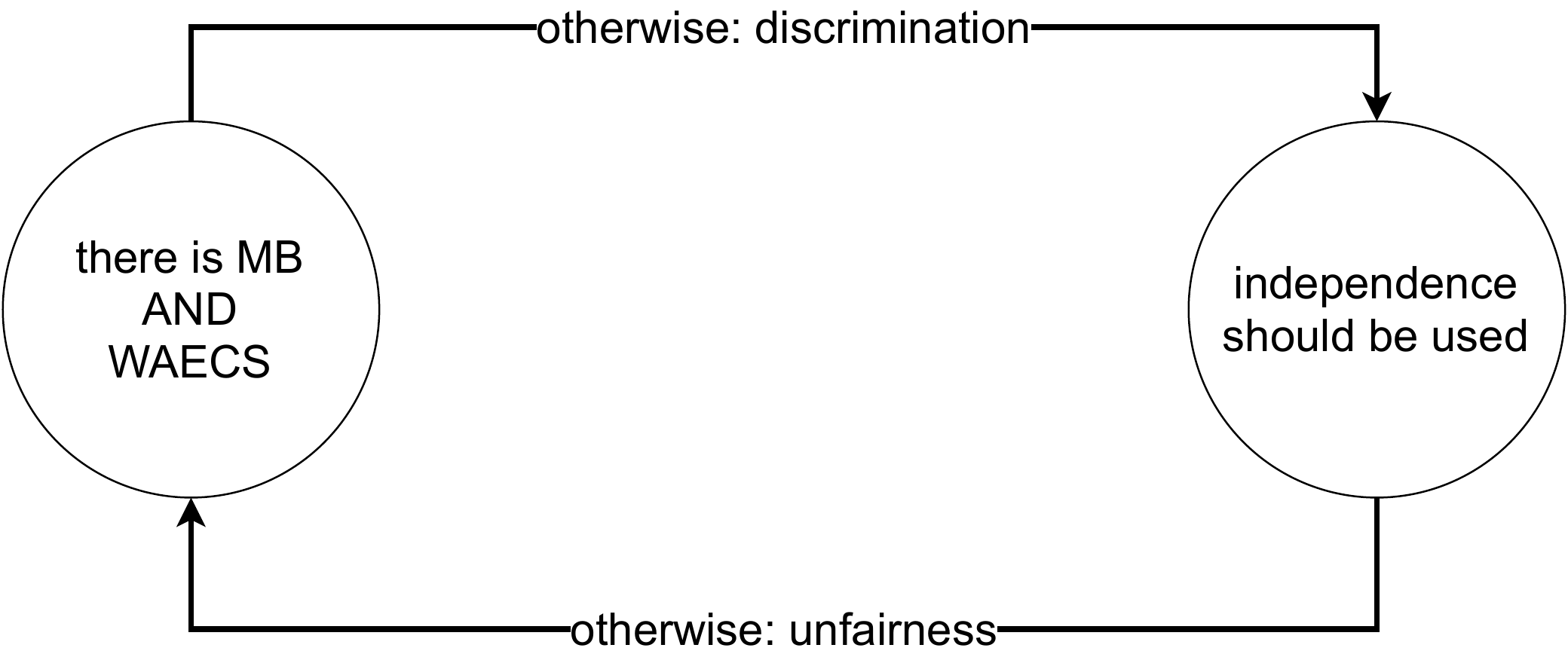}
    \caption{\citeauthor{friedler2016possibility}'s rules for choosing independence.}
    \label{fig:friedler-rules}
\end{figure}

Figure \ref{fig:friedler-rules} shows both rules and the implications of not following them.\footnote{
We can find similar interpretations of \citeauthor{friedler2016possibility}'s rules in the literature, e.g. in \cite{yeom2021avoiding, binns2020apparent, mitchell2021algorithmic}.
While there are subtle differences between their interpretations and ours (which we lack the space to discuss here in detail), we will show in \Cref{sec:counterexamples} that whichever interpretation of \citeauthor{friedler2016possibility}'s rules is chosen still falls short of justifying statistical parity as a general rule.
}

Note that \texttt{IFF x, THEN y} is equivalent to saying \texttt{IF x, THEN y AND IF y THEN x}.
We can therefore merge \Cref{prop:mb-waecs-independence} and \Cref{prop:independence-mb-waecs} as follows:

\begin{proposition}\label{prop:iff-mb-waecs-independence}
\texttt{IFF (there is \gls{MB} AND \gls{WAECS}),\\
THEN independence should be used}.
\end{proposition}

To illustrate this rule, let us consider the hiring example introduced in \Cref{sec:framework}.
We first assume that if we split the pool of candidates into demographic groups, then all groups are on average equal in their qualifications, e.g., their ability to sell the hiring company's product.
Second, we assume that standardized test scores shown on the candidates' CVs distort their actual qualifications.
To illustrate such a case of \gls{MB}, \citeauthor{friedler2016possibility} recite research, which calls into question the validity of standardized tests to assess verbal aptitude across different racial groups \cite{santelices2010unfair}.
In this case, independence should be enforced as both \gls{WAECS} and \gls{MB} are given.

Note though that the given reasons for enforcing or not enforcing independence ("fairness" and "non-discrimination") are by definition purely mathematical, not moral.
However, the terms still imply philosophical meaning.
\citeauthor{lipton2018troubling} \cite{lipton2018troubling} refer to the naming of technical concepts with terms that are colloquially used as \textit{suggestive definitions}:
These terms carry meaning in day-to-day life and therefore imply that our intuitive understanding of these terms in some way aligns with their technical definition.
This raises the question of whether these terms aim at not only providing a purely statistical reasoning about independence, but also a moral one.
As demonstrated above, if we interpret the paper literally, then it only provides mathematical justifications for the introduced rule.
However, since the \gls{DS} includes predictions that are used to make potentially life-altering decisions, it is hard to see how the rule could only be concerned with the mathematical validity of the predictions without any further moral considerations.
In the next section, we will therefore consider the natural extension of \Cref{prop:iff-mb-waecs-independence}:
We will view it as providing moral reasons for/against enforcing independence.\footnote{
\citeauthor{binns2020apparent}'s interpretation in \cite{binns2020apparent} considers this moralized version.
}
In doing so, we must consider all spaces introduced in \Cref{sec:extension}.
The moral rules that we introduce therefore include not only \gls{WAECS} and \gls{MB}, but also \gls{WAEPS} and \gls{LB}.

%% file: chapters/04_enhance_wae.tex
\section{Dealing with \gls{LB}}\label{sec:moral-rules}

As we have shown, the literal interpretation of the \gls{WAE} worldview states that all groups are actually the same in the \gls{CS}, even though it may look differently when taking measurements.
We should, however, consider what follows from assuming the \gls{WAE} worldview at the \gls{PS} level, while allowing for the existence of \gls{LB}.
Clearly, the result may be that we are no longer all equal in the \gls{CS} since \gls{LB} may affect members of different groups in different ways.

\subsection{Two Ways of Dealing With \gls{LB} in Distilled Rules}

We will now discuss how \gls{LB} may be included in the rules proposed by \citeauthor{friedler2016possibility}
The first way of dealing with \gls{LB} in \Cref{prop:iff-mb-waecs-independence} is to preserve the rule and simply apply it the same way when there is \gls{LB}.
This approach does not consider morality and instead prioritizes \citeauthor{friedler2016possibility}'s \cite{friedler2016possibility} original reasoning about mathematical properties.
These properties are what "fairness" (in its moral sense) is ultimately reduced to when dealing with \gls{LB} this way.
The second way of including \gls{LB} in the rule is to extend it and make the argument a moral one.
For this, we extend the rule to the new type of bias, \gls{LB}, by arguing by analogy: one could reason that \gls{LB} -- in terms of its moral features -- is sufficiently similar to \gls{MB}, so that similar rules follow when there is \gls{LB} instead of \gls{MB} or both.
Indeed, the analogy between \gls{MB} and \gls{LB} will be stronger if we can identify a common moral principle that applies to both cases.

Let us now examine both ways for dealing with \gls{LB} in turn.

\subsubsection{The mathematical extension}

Under this extension, we simply apply the rules that we distilled from \citeauthor{friedler2016possibility}'s framework, independently of whether there is \gls{LB} or not.
\Cref{prop:iff-mb-waecs-independence} states that \texttt{IFF (there is \gls{MB} AND \gls{WAECS}), THEN independence should be used}.
In order to see how this rule plays out when there is \gls{LB}, we consider the case where \gls{WAEPS} and inequalities in the \gls{CS} are caused by \gls{LB}.
This means that by hypothesis \texttt{NOT \gls{WAECS}}.
It follows from \Cref{prop:iff-mb-waecs-independence} that independence should not be used in this case.
The reasoning consistent with \citeauthor{friedler2016possibility} would be:
If \gls{WAECS} is not assumed, enforcing independence violates \citeauthor{friedler2016possibility}'s mathematical fairness property (that individuals close in the \gls{CS} should be close in the \gls{DS}) -- independently of whether they are equal in the \gls{PS} or not.
This line of argument may, however, leave scholars that are interested in fairness (in its moral sense, not \citeauthor{friedler2016possibility}'s statistical concept) dissatisfied.

\subsubsection{The moral extension}

In \cite{friedler2016possibility}, the argument in favor of fairness does not provide an explicit moral grounding to define fairness the way it is defined.
This definition of fairness is explicitly inspired by \citeauthor{dwork2012fairness}'s \cite{dwork2012fairness} notion, which relies on similarity in the \gls{CS} to determine what makes a decision \textit{fair}.
However, it may be objected that similarities in the \gls{CS} under the influence of injustice do not provide a suitable reference point to define what is a fair prediction or decision \cite{binns2020apparent}.

To see this, let us consider the case of credit lending described by Reuben Binns in \cite{binns2020apparent}.
We assume that we observe in the \gls{OS} that women have historically been less likely to repay their loans than men.
\citeauthor{binns2020apparent} provides two possible reasons for this.
The first is that credit lenders may be more lenient towards men, allowing them delays in their repayment.
This can be interpreted as \gls{MB}: Women and men who are equally "good" at repaying their loans end up with different credit histories (which are proxies for their repayment ability) as men are treated preferentially.
A second reason could be that women are more likely to be single parents, which makes the repayment of credits more difficult.
This is an example of \gls{LB}: Due to societal structures, the average woman is less likely to repay her loan.

The question is then whether differences in the \gls{OS} between men and women (i.e., their credit histories) should be considered to be just.
\citeauthor{binns2020apparent} argues that if they are not considered to be just (e.g., because they are unjustly caused by gendered social structures), then enforcing independence should be considered. 

We may argue that both of the given reasons for the difference in the payment history (the existence of \gls{MB} and \gls{LB}) are unjust.
First, we could say that it is \textit{unjust} (and not merely, inefficient) for a decision about the individual to be taken when \gls{MB} exists.
Second, we may argue that it is unjust (and not merely, inefficient) for a decision about the individual to be taken when \gls{LB} exists.
This suggests that there is a common moral \textit{reason} for why decisions influenced by \gls{MB} or \gls{LB} are unfair.
We will now attempt to identify this common moral cause.

For this, we will follow the philosophical analysis, which is explicitly invoked by \citeauthor{binns2020apparent}.
This analysis will explain why both \gls{MB} and \gls{LB} cause unfair predictions and decisions.
It appeals to the question of \textit{responsibility} which asks whether individuals are responsible for predictions others make about them that impact their well-being.
This responsibility may fail to obtain either because the measurements that decision-makers have (the \gls{OS}) do not reflect people’s choices (i.e., people are not responsible for \gls{MB}) or because the construct that is measured (the \gls{CS}) does not reflect people’s choices (or both).
Thus, one should ask whether the \gls{OS} or the \gls{CS} is the result of \textit{choices} or of \textit{circumstances} individuals cannot control.\footnote{
In addition, in the case of \gls{MB} (but not \gls{LB}), a distinct moral argument can be given as to why a decision affected by \gls{MB} is unfair, which is based on merit.
The conventional view of merit is that it is based on what people do \cite{olsaretti2006justice, miller1979social}, for example, their contribution to society.
Suppose that the actual contribution to society of two people, A and B, is equal, that is A and B have the same \gls{CS} features.
However, \gls{MB} exists, so A is perceived to contribute more and, consequently, A receives a benefit that is denied to B.
This is intuitively unfair since A does not deserve a favorable treatment compared to B (A has not contributed more to society than B).
}
When \gls{MB} exists, the individuals who are judged in a biased manner do not control the bias and are not morally responsible for it.
Intuitively, it is unfair that individuals are imposed costs due to factors for which they are not responsible.
We can thus describe the moral view behind proposition \Cref{prop:iff-mb-waecs-independence} in the following way: if \gls{WAECS} and \gls{MB} is assumed, then differences in the \gls{DS} are unjust on responsibility grounds.

Similarly, one may consider people’s actual abilities and behaviors as responses to the peculiar circumstances in which people happen to be born and grow up (which are not up to them).
Thus, people are not responsible for the influence of those peculiar circumstances, i.e., for their \gls{LB}.
It follows that people should not benefit or get harmed or, more generally, be treated differently because of \gls{LB}, \textit{which manifests itself in the \gls{CS}}. 
Hence, by parity of reasoning -- according to the above moral interpretation of \Cref{prop:iff-mb-waecs-independence} -- if we seek to eliminate the influence of \gls{MB} on the decision, we should also seek to eliminate the influence of \gls{LB} on the decision.

\paragraph{Conclusion of the moral extension.}

This suggests the following view as the natural extension of \Cref{prop:iff-mb-waecs-independence}, i.e., the view that \texttt{IFF (there is \gls{MB} AND \gls{WAECS}), THEN independence should be used}:
\begin{proposition}\label{prop:waeps-lb-mb-indepdence}
    \texttt{
    IFF \gls{WAEPS} AND (there is \gls{MB} OR\footnote{When using \texttt{OR}, we are referring to the logical operator $\lor$, which means that the statement is true if either one or both operands are true.} there is \gls{LB}), THEN independence should be used.
    }
\end{proposition}

\subsection{Not All \gls{LB} Should Be Corrected}\label{sec:just-unjust-lb}

Intuitively, \Cref{prop:waeps-lb-mb-indepdence} states that independence is called for not only to correct for \gls{MB}, but also to correct for \gls{LB}.
Notice that \Cref{prop:waeps-lb-mb-indepdence} is now arguably too broad in the inequalities it promises to correct for.
The moral argument for removing the influence of \gls{MB} was that it was neither morally neutral, nor merely inefficient, but actually \textit{unjust} as it does not reflect merit or responsibility.
Something similar, intuitively, must hold in this case. 
Namely, it is \textit{unjust} \gls{LB} that calls for some kind of correction or rectification.

\subsubsection{Distinction between \textit{just} and \textit{unjust} \gls{LB}}

Note that it is a logical possibility that indeed all \gls{LB}, \textit{as such}, is unjust.
If so, there is no distinction between \textit{unjust} \gls{LB} and \gls{LB} simpliciter.
This is the position that no one deserves the values in the CS which are influenced by any type of \gls{LB}.
It is, however, also possible to maintain a more nuanced view.
It is easy to show this by considering theories of justice that political philosophers \textit{actually} defend.
Different substantive theories of justice provide different (and often irreconcilable) criteria for evaluating the justice of social structures.
For example, institutional luck egalitarianism maintains that all inequalities (in the metric of what matters ultimately for justice, e.g. well-being) for which individuals are not responsible are unjust \cite{dworkin_what_1981, cohen_currency_1989}.
Unjust inequalities are the ones which could have been prevented or redressed through suitable and feasible institutional arrangements.
\citeauthor{rawls2001justice}'s theory of justice, on the other hand, maintains that inequalities reflecting people's unequal native endowments and motivations are just -- provided that (1) they are not influenced by the social class of birth and (2) emerge through institutions arranged in a way that delivers the greatest expectations of social primary goods to society's least advantaged members \cite{rawls2001justice}.

These theories (and many others) disagree when arguing about the justice of institutions.
The luck egalitarian one, for example, commits the government to do everything it can to level the playing field among individuals born with unequal natural endowments, as these inequalities are undeserved.
\citeauthor{rawls2001justice}'s view, however, approves of such inequalities if they boost productivity in a society where people at the bottom of the social pyramid are the ones to benefit the most from such productivity gains \cite{rawls2001justice}.
Yet these views also converge on many real world cases: for example, current US society is arguably very unjust according to both views.

From  the perspective of both luck egalitarianism and \citeauthor{rawls2001justice}'s theory of justice, many inequalities in the \gls{CS}, which are produced by \gls{LB}, actually reflect unjust social structures.
They are instances of \textit{unjust} \gls{LB}.
Notice that neither theory implies that all \gls{LB} is unjust \gls{LB}.
Let us consider the luck egalitarian view that only factors for which one can be held morally responsible justify inequality.
It may still be objected that the development of innate potential into realized abilities is not determined entirely by external circumstances that are matters of brute (good or bad) luck.
Apart from the influence of good and ill fortune, our realized abilities reflect our personal history, i.e., the choices we make, every day.
If human agency is not an illusion, we are (partially, at least) responsible for at least some of our choices.
Therefore, at least some \gls{LB} is not morally problematic.
It is therefore unclear why one should treat inequalities arising in the \gls{CS} as a result of such \gls{LB} on a par with inequalities arising in the \gls{OS} due to \gls{MB}.

\subsubsection{Just \gls{LB} at group-level}

In the discussion above, we are saying that not all \gls{LB} is necessarily unjust because some \gls{LB} might simply be caused by personal choices.
However, one may question whether it is possible for personal choices to cause unequal outcomes not only between individuals, but also at the group level, even if there were no inequalities in the \gls{PS}.
For example, two people with the same potential, with institutions that only let inequalities reflecting their individual choices exist, could still end up with different realized abilities.
Yet, on a group level, it seems improbable that one group justly has a statistical prevalence of individuals making one kind of choices, for which they can be held responsible, and another group justly has a statistical prevalence of individuals making a different kind of choice.
If that different prevalence exists and we assume \gls{WAEPS}, certainly there must be something causing the group inequality for which individuals cannot and shall not be held responsible.
The question is, therefore, whether \textit{just} \gls{LB} is actually possible on a group-level (which is the relevant level when talking about enforcing independence).

In reply, there are at least two ways of showing the possibility of just \gls{LB}.
First, we consider moral views that differ from luck egalitarianism.
Consider the view that the influence of parents who read bedtime stories to their children and in this way cause unequal IQ, i.e., differences in the \gls{OS}, is never unjust \cite{mason2006levelling}, a view that contradicts luck egalitarianism \cite{segall2011if}.
If that view is correct, inequalities that have been created by reading bedtime stories to children are not unjust even if it so happens that, for historical and cultural reasons, reading bedtime stories to children is more habitual in certain cultures than others.

Second, just \gls{LB} may exist under a luck egalitarian view if we do not assume \gls{WAEPS}.
Let us consider the case of a genetic disease which is more common in a specific population due to the founder effect.
Spending more resources on the detection and treatment of this disease for the population that is most affected (e.g., more medical check-ups, financial support for therapies) is by definition a form of \gls{LB}.
This is because if we assumed \gls{WAEPS}, investing more in the detection and treatment of this disease for people in the particularly affected population is the sort of circumstance that would create inequalities in the \gls{CS}, i.e. the rest of the population would on average suffer more from the effects of the disease.
However, in a world in which we are not all equal in the \gls{PS} (as is the case with this genetic disease), the increased spending on the more affected population could plausibly be seen as a case of \textit{just} \gls{LB}.
Under a luck egalitarian view of justice, for example, this \gls{LB} would clearly be just because it mitigates an inequality in the \gls{CS} for which individuals are not morally responsible (i.e., that without intervention, one group is more likely to suffer from the disease).%
\footnote{
Understanding this difference between just and unjust \gls{LB} can help us understand how our extension of \citeauthor{friedler2016possibility}'s framework relates to \citeauthor{mitchell2021algorithmic}'s \cite{mitchell2021algorithmic} proposition to differentiate between two notions of biased data: "statistical bias" and "societal bias".
"Statistical bias" occurs between what they refer to as the "world as it is" (i.e., the \gls{CS}) and the "world according to data" (i.e., the \gls{OS}).
It is thus simply another term for \gls{MB}.
"Societal bias" is introduced from the "world as it should and could be" to the "world as it is".
One may think that the "world as it should and could be" is equivalent to our \gls{PS} and that "societal bias" is thus equivalent to \gls{LB}.
However, the \gls{PS} is a purely descriptive notion while the "world as it should and could be", while never defined, suggests a normative concept: To define what the world ought to look like, a philosophical concept is needed.
Such a philosophical concept might morally require the introduction of just \gls{LB} that ensures that the "world as it is" reflects the "world as it should and could be".
In this case, our reading of \citeauthor{mitchell2021algorithmic} is that there is no "societal bias" whereas our model would note that there is \gls{LB}, but that this \gls{LB} is just.
Our extension thus gives us the tools to describe the existence \gls{LB} without yet making normative judgments about it.
Such normative judgments are only required when differentiating just from unjust \gls{LB}.
}

When \gls{LB} exists, but is not unjust, the moral reason for enforcing independence no longer holds.
In other words, the decision whether independence should be used depends not only on facts but also on values when \gls{LB} exists.
According to luck egalitarianism, independence is not required if the following two requirements are fulfilled:
(1) unequal decisions (i.e., inequality in the \gls{DS}) emerge purely as a result of unbiased observations (i.e., \gls{WYSIWYG}) of the features that we \textit{want} to base the decision on (i.e., \gls{CS}) and
(2) these features are unequally distributed (in spite of equality of potential in the \gls{PS}) simply as a result of choices for which individuals can be considered \textit{fully responsible}.
According to other theories of justice, independence is not required if the features in question emerge as a result of processes such as reading bedtime stories to one's children.
Whether independence should be enforced or not thus depends on one's view of what \gls{LB} is just.

\subsection{Extended Rules: Final Formulation}

We have argued that if we consider \gls{LB}, we should consider a moralized version of the relation between \gls{LB} and independence (i.e., not all forms of \gls{LB} require to be corrected).
As we have shown, the theory of justice one adheres to determines one's judgment about the justness of the \gls{LB}.
This in turn determines one's judgment as to whether \gls{LB} provides a reason to enforce independence or not.
In conclusion: the most charitable interpretation of the extension of \Cref{prop:iff-mb-waecs-independence} to include \gls{LB} is not \Cref{prop:waeps-lb-mb-indepdence}, but rather the following more nuanced view:

\begin{proposition}\label{prop:waeps-unjust-lb-mb-indepdence}
    \texttt{
    IFF \gls{WAEPS} AND (there is \gls{MB} OR there is unjust \gls{LB}), THEN independence should be used.
    }
\end{proposition}

In what follows, we will consider \Cref{prop:waeps-unjust-lb-mb-indepdence} as a natural extension of \Cref{prop:iff-mb-waecs-independence}.
However, we want to focus our attention on the implications of \Cref{prop:waeps-unjust-lb-mb-indepdence} in the scenario in which \gls{MB} does not exist in order to simplify our discussion somewhat.
We will refer to the remaining underlying assumptions as the \gls{WAEBI} worldview.

\theoremstyle{definition}
\begin{definition}[\gls{WAEBI} worldview]\label{def:waebi}
The \gls{WAEBI} worldview subsumes the following assumptions:
\begin{enumerate}
    \item \gls{WAEPS} and
    \item there is unjust \gls{LB}.\footnote{Note that this second assumption logically leads to the assumption of unjust inequalities in the \gls{CS}.}
\end{enumerate}
\end{definition}

\Cref{prop:waebi-indepdence} follows from \Cref{prop:waeps-unjust-lb-mb-indepdence} if it is assumed that there is no \gls{MB}.

\begin{proposition}\label{prop:waebi-indepdence}
    \texttt{IFF \gls{WAEBI}, THEN independence should be used.}
\end{proposition}

In the hiring example, \gls{WAEBI} is given if, for example, (1) at birth, all demographic groups have the same average potential to become hireworthy sales people, but (2) this equality is lost because one group has, on average, less money and is thus more likely to have to work odd jobs instead of doing unpaid internships.
The resulting differences in the candidates' qualifications and proxies like their CV are thus considered to be unjust and should be corrected by enforcing independence.

%% file: chapters/05_arguments.tex
\section{Two Counterexamples Against Extended Rules}\label{sec:counterexamples}

We will now examine if \Cref{prop:waeps-unjust-lb-mb-indepdence} holds up as a general rule for when to enforce independence.
Clearly, it does not represent a general rule if we can find cases in which this rule does not apply.
The goal of this section is to see if we can find such cases.

As stated before, we focus our search for counterexamples on \Cref{prop:waebi-indepdence}, which simplifies the discussion of \Cref{prop:waeps-unjust-lb-mb-indepdence}.
\Cref{prop:waebi-indepdence} is the claim that \texttt{IFF \gls{WAEBI}, THEN independence should be used}.
When we assume that there is no \gls{MB}, \Cref{prop:waebi-indepdence} is true if and only if \Cref{prop:waeps-unjust-lb-mb-indepdence} is true.
Thus, we can simplify the analysis of \Cref{prop:waeps-unjust-lb-mb-indepdence} somewhat by focusing on \Cref{prop:waebi-indepdence}, assuming that no \gls{MB} exists.
We shall proceed in a logical fashion by investigating the two parts of the biconditional in turn:
\begin{enumerate}
    \item \texttt{IF {WAEBI}, THEN independence should be used} and\label{enum:arg-1}
    \item \texttt{IF independence should be used, THEN {WAEBI}}\label{enum:arg-2}
\end{enumerate}
Notice that rule \ref{enum:arg-2} is informative, even though we already know that it is incorrect because independence should also be used when \gls{MB} exists, even if \gls{WAEBI} is not assumed.
The reason why it is still informative is that we will show that the claim is incorrect -- independently of \gls{MB}.

We argue against both rule \ref{enum:arg-1} and \ref{enum:arg-2} by counterexample.
The counterexamples will show not only that \Cref{prop:waebi-indepdence} is incorrect, but also that \Cref{prop:waeps-unjust-lb-mb-indepdence} is since the counterexamples do not involve \gls{MB}.
Thus, we refer to \Cref{prop:waebi-indepdence}, so that we can bracket the issue of \gls{MB} and avoid distractions.

\subsection{Counterexample Against Rule 1: IF \gls{WAEBI}, THEN independence should be used}

It is now time to offer a convincing counterexample to the view that (absent \gls{MB}) independence should be enforced if \gls{WAEBI}.
For this, recall the definition of \gls{WAEBI}, \Cref{def:waebi}, which states that this worldview assumes \gls{WAEPS} and unjust \gls{LB}.
The counterexample is a case in which the \gls{WAEBI} assumptions are all satisfied, yet independence is not required.
One can build a counterexample as follows:
\begin{enumerate}
    \item \label{enum:counterexample-1-1} First, let us suppose that there is a specific severe congenital disorder that is very painful and drastically reduces the individual's life expectancy.
    We will refer to this specific severe congenital disorder as \textit{SCD}.
    Let us further assume that (probably contrary to fact) all individuals are generally equally at risk of being born with \textit{SCD}.
    \gls{WAEPS} is therefore satisfied.
    \item \label{enum:counterexample-1-2} Second, let us suppose that -- while the risk for being born with \textit{SCD} is generally the same for all individuals -- this risk is notably increased when the mother breathes in dangerous pollutants during pregnancy.
    Assume now that mothers in one group, e.g. the \textit{green} group, are more likely to live in neighborhoods close to chemical factories that emit dangerous pollutants.
    Individuals in the green group are thus more likely to be exposed to the risk of developing \textit{SCD}.
    We shall suppose that this unequal exposure is produced by huge and uncontroversial injustices in society.
    \textit{Green} mothers might, for example, be more likely to live in poverty because of direct discrimination against them, which makes their opportunities for all sorts of job worse than those of the \textit{orange} group.
    For this reason, they cannot afford moving and have to live in poor neighborhoods plagued by dangerous pollutants.
    (This case plausibly counts as injustice even according to more moderate forms of egalitarianism than luck egalitarianism.)
    Hence, there is unjust \gls{LB}.
    \item \label{enum:counterexample-1-3} Third, as a result of \ref{enum:counterexample-1-2}, members of the \textit{green} group are more likely to suffer from \textit{SCD} than members of the \textit{orange} group.
    Thus, there is an unjust inequality in the \gls{CS}.
    We shall suppose that whether a patient suffers from \textit{SCD} is a clear cut, binary condition, i.e., either someone does, or does not.
    There are no intermediate stages.
\end{enumerate}

Suppose that a very expensive therapy is developed, which cures people with \textit{SCD} but causes recurrent migraine (with moderate frequency, let us say, once per month).
As \textit{SCD} has bad consequences for the individual (pain, drastically shortened life), we shall assume that the benefits of the therapy outweigh its costs.
Suppose now that the therapy only works if it affects fetal development.
Thus, in order to avoid the disease for the future individual, it is the mother that has to be treated before the illness is fully manifested in the child.

Suppose that machine learning specialists develop a perfect accuracy predictor to determine, based on a non-invasive clinical examination, whether the fetus will be ill.
(This may be impossible in practice. However, in a philosophical argument, we can test the theory also with hypothetical examples. The challenge is to explain what could be morally wrong with the independence-fulfilling predictor. Notice also that in the clinical setting one can already make high accuracy predictions. With close to perfect accuracy, people often act and reason as if the accuracy was perfect.)
Since the predictor is perfectly accurate, it will predict \textit{SCD} at a higher rate for the green than for the orange patients.
As a result, green patients will receive the therapy more often than orange patients do, which violates independence.

We will now argue that this perfect accuracy predictor is perfectly just.
The argument we present is very robust because it is coherent with ethical views that sometimes pull in different directions and, intuitively, it is difficult to make the case that the argument is wrong.
Indeed, it should be so obvious that the predictor is fair, that it would be counted against any view entailing the opposite for this case, that it cannot align with this result.
The decision of the perfect predictor is perfectly fair because no individual has a claim against the distribution based on it.
By "no individual has a claim", we do not mean that some individual may have a \textit{prima facie} claim that a different decision should be taken, which is then overridden by the claims of others.
We also do not mean that some individual has a claim that holds \textit{prima facie}, but that is defeated by some substantive view of justice, which the individual, if reasonable or endowed with moral sensibility, should respect (even if it is not in the individual's own interest to respect it).
What we actually mean by "no individual has a claim" is the much more radical claim that the individual has no claim against the perfectly accurate distribution (in this case), not even a \textit{pro tanto} or a \textit{prima facie} claim.

To see why no individual has a claim against the perfect accuracy distribution, consider that no individual, faced with the decision by a perfect predictor, can point to an alternative distribution that they have any reason to prefer.
This clearly is the case in the example.\footnote{It may be objected that this is a very peculiar example, and that not all cases involving perfect accuracy predictors are relevantly similar. That is probably correct. However, one case is all it takes to generate a counterexample that falsifies a general claim about when independence should be used.}
For, first, each individual person who will develop \textit{SCD} is better off with a decision based on the correct prediction because the individual is certain to receive the cure, which is the preferable outcome despite the side effects.
Second, every individual who will not develop \textit{SCD} is better off without the therapy because the individual is certain to not need the cure.
Not receiving the therapy is thus the preferable outcome as it avoids the side effects.
As a consequence, no one has a claim to a different decision.

Moreover, any departure from the perfect accuracy predictor makes someone worse off and no one better off.
When the features in the \gls{CS} are not equally distributed between the two groups (i.e., \textit{green} and \textit{orange}, in this case), enforcing independence sacrifices some accuracy.
Suppose that this sacrifice amounts to a single wrong diagnosis.
That means: either someone who will actually develop \textit{SCD} will not receive the cure, or someone who will not develop \textit{SCD} will receive the cure.
Either way, the choice to enforce independence will cause harm to at least one individual, which gives that individual a claim \textit{against} independence.
It seems that this is one rare case in which one view of what is fair is truly robust because, besides maximizing utility, no individual has a claim against the perfect accuracy predictor, even if it violates independence.
Furthermore, if independence is enforced in this case even though it causes inaccuracy, there will be at least one individual who has a moral claim against independence being enforced.
This claim entails that enforcing independence is morally wrong because it is not defeated by any claim \textit{in favor of} enforcing independence.
The question of comparing the relative urgency or strength of moral claims does not even arise.

Our argument here is not merely that independence in this case involves a loss of accuracy (and thus utility) and that, simply for that reason, is the morally wrong choice in this case.
While it is correct that there is a conflict between independence and accuracy in this case, our argument is much stronger than the usual utilitarian argument.
The usual utilitarian argument points out that enforcing independence causes a loss of \textit{aggregate} utility \cite{corbett2017algorithmic}.
This argument also focuses on utility, but it considers it from the perspective of each and every individual involved in the decision.
A utilitarian argument would object that enforcing independence causes a utility loss in the aggregate and that for that reason it should not be done.
However, such an argument also requires that, in order to reach the utilitarian maximum, some people are made worse off for the benefit of other people.
The utilitarian view is that this is always morally right when the sum of utility is maximized.
Many people find this view objectionable (e.g., \cite{rawls2001justice}).
The objection against the utilitarian is that it does not respect the \textit{separateness of persons} \cite{rawls2001justice}.
Our argument against independence does not imply the utilitarian conclusion, so it is is not vulnerable to this objection.%
\footnote {
Our argument is a contractualist one, not a utilitarian one \cite{scanlon2000we}.
Our thesis that the perfect accuracy predictor is just is so robust because it is independently supported by contractualism and utilitarianism.
}

Summing up, it is not true that \texttt{IF \gls{WAEBI}, THEN independence should be used}.
In this case, \gls{WAEBI} is clearly satisfied (by hypothesis), and yet independence should not be used. 

\subsection{Counterexample Against Rule 2:  IF independence should be used, THEN {WAEBI}}

Now let us turn to the other direction of the biconditional, which is the idea that \texttt{IF independence should be used, THEN {WAEBI}} is assumed.
A counterexample to this would be a case in which independence seems intuitively fair or called for, yet {WAEBI} conditions are not satisfied.
Unfortunately, this example is not as robust as the first one is.
The example itself is inspired by a fairness theory for machine learning, which is based on economic and political theories of equality of opportunity and which provides indications for when independence should be used \cite{heidari2019moral}.
We do not rely on this theory, as we find that a strong case can be made for the conclusion on intuitive grounds.  

We consider the design of an algorithmic decision system deployed after natural disasters.
This decision system is tasked with determining where drones should be sent in order to attempt to rescue civilians from drowning after their houses and streets have been flooded.
Data scientists train a machine learning model to decide where to send the drones in such cases.
The initial goal is to simply maximize the number of lives saved.

Let us assume that there is a flooding which affects a city with its surrounding suburbs.
While the city is densely populated, the suburbs are not.
We can split the population into two demographic groups: the \textit{green} and the \textit{orange} group.
It turns out that the \textit{orange} group tends to live in the city and the \textit{green} group tends to live in the suburbs.
Because of the difference in population density between the city and the suburbs, a drone that is sent to the city has a much higher probability of resulting in a successful rescue.
Hence, the utility-maximizing model is more likely to send drones to the densely populated city than to the suburbs -- it diverts resources to the suburbs only when a large proportion in the cities has been saved.
As a consequence, the probability to be saved is much higher (say, ten times higher) if you are \textit{orange}.
This means that members of the \textit{green} population are very unlikely to be rescued.%
\footnote{
Clearly we assume here that data scientists cannot reach, or even approximate, a perfect accuracy predictor.
This implies that when the algorithm predicts that a person will be saved by sending a helicopter to coordinates $X$, $Y$, $Z$, it is not always the case that someone will get rescued, particularly in the suburbs.
}
We maintain that in this case independence is morally required.
The reasoning is the following:
Every individual equally needs to be saved, independently of where they live, and no one should be held morally responsible for failing to live in a relatively densely populated area, for matters of life and death.
Thus, in a sense, everyone equally deserves to be saved.
If everyone equally deserves to be rescued, everyone should have the same prospects of being rescued, independently of their sex, race, or any other sensitive attribute.
If so, any inequality in the probability of rescue associated with membership to a group is morally problematic, for it cannot be justified based on merit, or need, or responsibility.

It may be objected that there is a clear moral reason to prioritize saving urban individuals, namely that this will maximize the total number of lives rescued (and we ought to maximize this value).
However, notice again, that this is a utilitarian, maximizing argument.
Most moral problems of fairness in machine learning, or at least most \textit{morally deep} problems, emerge because there is a conflict between maximizing utility and fairness (in its moral sense) defined in a way that is independent from it.
Hence, in a sense, the fact that the fairness intuition conflicts with a utilitarian assessment of what should be done is to be expected for an authentic (non-utilitarian) moral intuition for fairness.
Arguably, the best way to take the utilitarian intuition -- that there is a (moral) reason to send drones predominantly into the densely populated areas -- into account is by viewing it as a consideration of efficiency that an ethically sound procedure should balance with considerations of distributive justice.
That is, the \textit{all-things-considered} morally desirable algorithm will neither be one that maximizes utility, nor one that achieves independence fully, but rather something in between, that will compromise utility, to some extent, but also achieve a more balanced rescue of the two populations.
We then conclude that this is a case in which independence should be used due to a concern with fairness (in its moral sense).
(Even if, let us grant the objection, not fully achieved as fairness needs to be balanced with efficiency.) 

Having argued that independence should be used, this amounts to the counterexample we are looking for if we show that the moral case for independence is independent of the {WAEBI} conditions.
For building the stronger possible case, we shall suppose that \textit{every single one} of the conditions that jointly define the \gls{WAEBI} worldview is false.

First, we do not assume that \gls{WAEPS}, that is, people are born with a disposition to live in cities or suburbs.
For example, some people live in the city just because they are born there, even if it is not true of everyone.

Second, it is not the case that \gls{LB} exists, or at least, the plausibility of the conclusion about fairness does not depend on the existence of \gls{LB}.
We may consider, for the sake of the argument, a society in which people are not pressured to live in cities.
The case for rescuing the people in the suburbs is as strong in a society in which people are not pressured to live in cities, as it is in one in which they are pressured to do so (among other things, by the perception that their lives have less value in the eyes of rescue drones if they remain in less densely populated areas).
So the conclusion does not depend on the existence of \gls{LB}.

Third, we may as well suppose that there is \gls{LB}, but the \gls{LB} is not \textit{unjust}.
For example, people end up living in suburbs and cities (and different groups, e.g., \textit{green} and \textit{orange}, have different propensities to do so), but this is not in itself unjust or the result of injustice in society.
Schelling's model of segregation shows that a mild preference for living among members of the same group will over time lead to segregation \cite{schelling1971dynamic}.
For this example, we assume that the \textit{green} and \textit{orange} population have a slight preference for members of their own group and that this preference is innate and not caused by injustices.
Over time the two groups have segregated to some extent, so that the majority of the people living in the city happens to be \textit{orange} and the majority of the people living in the suburbs is \textit{green}.\footnote{
For examples of \textit{unjust} causes of segregation see \cite{power1983apartheid, silver1991racial, frey1979central, charles2003dynamics, archer2019housing}.
}

In conclusion, we have identified a case in which independence should (plausibly) be used.
And yet, in this case, the conditions realizing the \gls{WAEBI} worldview are not satisfied.
This counts as a counterexample to the claim that \texttt{IF independence should be used, THEN \gls{WAEBI}}, and concludes our rebuttal of the biconditional claim \Cref{prop:waebi-indepdence}.
Since neither example depends on the existence of \gls{MB}, the arguments also disprove \Cref{prop:waeps-unjust-lb-mb-indepdence}.

%% file: chapters/06_conclusion.tex
\section{Conclusion}\label{sec:conclusion}

In this paper, we have analyzed one argument that can be given in support of enforcing independence in a machine learning model, found in the recent machine learning literature.
This argument claims that one shall enforce independence (i.e., use it as a fairness constraint of the model) if (and only if) \acrfull{WAE} and there is \acrfull{MB}.
We have introduced the concept of \acrfull{LB} as a type of bias, which influences how the potential an individual is born with develops into realized abilities.
This bias can be distinguished from the \gls{MB} proposed by \citeauthor{friedler2016possibility} (They call this type of bias "structural bias".)
This shows that the \gls{WAE} view as stated in the literature is incomplete as demographic groups can be equal not only with respect to their realized abilities but also their potential.

We have identified two possible extensions of the argument presented in the literature, which are relevant when inequalities are generated by \gls{LB}.
We argue that the most (morally) plausible extension is the view that one should enforce independence if (and only if) there is \gls{MB} or if \acrfull{WAEBI}.
In other words, it seems like independence could be justified when taking on the \gls{WAEBI} worldview, which assumes that socio-demographic groups have similar innate potential at birth, but unjust \gls{LB} leads to differences in their realized abilities.

Unfortunately, we found two powerful counterexamples to this ideally simple view:
the first clearly showed that unjust \gls{LB} does not always morally require enforcing independence; the second made it plausible that (even in the absence of \gls{MB}) injustice is not required for the use of independence to be morally justified.

The relatively simple and morally plausible proposition linking \gls{WAEBI} and independence we analyzed here is thus not universally true.
One may object to the first counterexample, saying that it presents a case where what is being distributed is not (uniformly) beneficial, that is, the treatment would be a net harm for many of the subjects.
However, it is true of many cases discussed in the algorithmic fairness debate that what is being distributed is not uniformly beneficial: arguably, even being admitted to a university that is too demanding for one’s skills might be harmful and being released on parole may not be beneficial for the parolee who in fact reoffends and re-enters prison with a worse criminal record.
One may further object that in the first counterexample, considering efficiency alone would produce a fair outcome.
We argue that if there are cases in which the efficient solution is clearly and intuitively considered the fair solution, we need a philosophical theory that can explain why this is in fact the case.
Our argument thus reveals that a promising line of research may be built by judging the morality of fairness metrics not only based on the question of what causes differences in predictions, but also based on how these predictions distribute utility.